\documentclass[12pt,dvips]{article}

\usepackage{cite,epsfig}
\usepackage{color}

\def\lsim{\:\raisebox{-0.5ex}{$\stackrel{\textstyle<}{\sim}$}\:}
\def\gsim{\:\raisebox{-0.5ex}{$\stackrel{\textstyle>}{\sim}$}\:}

\begin{document}

\begin{flushright}
MAN/HEP/2008/4\\
arXiv:0802.3197 \\
February 2008
\end{flushright}

\begin{center}
{\bf {\LARGE $\Upsilon\to\gamma A_1$ in the NMSSM at large 
{\boldmath $\tan\beta$}}}
\end{center}

\bigskip\bigskip

\begin{center}
{\large Robert N. Hodgkinson}\\[3mm]
{\em School of Physics and Astronomy, University of Manchester}\\
{\em Manchester M13 9PL, United Kingdom}
\end{center}

\vspace{2cm}

\centerline{\bf ABSTRACT}
\noindent

We investigate the effects of the radiatively-generated 
$\tan\beta$-enhanced Higgs-singlet Yukawa couplings on the decay
$\Upsilon\to \gamma A_1$ in the NMSSM, where $A_1$ is the lightest
CP-odd scalar. This radiative coupling is found to dominate in
the case of a highly singlet Higgs pseudoscalar. The branching ratio
for the production of such a particle is shown to be within a few
orders of magnitude of current experimental constraints across a
significant region of parameter space. This represents a potentially
observable signal for experiments at present $B$-factories.

\newpage


The Next to Minimal Supersymmetric extension of the Standard Model
(NMSSM) is a well-motivated model of electroweak symmetry breaking
which resolves both the hierarchy problem of the Standard Model (SM)
and $\mu$ problem of the Minimal Supersymmetric extension of the
Standard Model (MSSM) in a natural way \cite{NMSSM}. The $\mu$ 
parameter of the MSSM is replaced with an additional gauge singlet 
Higgs superfield $\hat S$ and an effective doublet mixing term 
$\mu_{\rm eff}$ is generated when the singlet field acquires a
vacuum expectation value (VEV). It has long been known that the
NMSSM suffers from the formation of electroweak scale cosmic
domain walls\cite{DomainWalls}, although mechanisms to resolve this
problem have been suggested, e.g.~\cite{DWResolution}. In the NMSSM,
all parameters are naturally predicted to be of the order the
SUSY-breaking scale $M_{\rm SUSY}$.

The Higgs sector of the NMSSM may be derived from the
superpotential of the model, given by
\begin{equation}
{\mathcal W}_{\rm Higgs} =
\lambda \hat S \hat H_1 \hat H_2 + \kappa \hat S^3,
\end{equation}
where $\hat H_1 (\hat H_2)$ is the doublet Higgs superfield which gives
masses to the down-type quarks and leptons (up-type quarks). The
corresponding soft SUSY-breaking terms are given by
\begin{equation}
{\mathcal L}_{\rm Higgs}^{\rm soft} = 
\lambda A_\lambda S \Phi_1 \Phi_2 +  \kappa A_\kappa S^3,
\end{equation} where $\Phi_{1,2}$ and $S$ are the scalar components of
$\hat H_{1,2}$ and $\hat S$ respectively. At tree level only two further
parameters are required, the ratio of doublet VEVs $\tan\beta=\frac{v_2}{v_1}$
and the effective doublet mixing parameter
$\mu_{\rm eff}=\frac{\lambda v_S}{\sqrt{2}}$. Radiative
corrections due to the quarks and scalar quarks of the third
generation must also be included in order to raise the mass of
the SM-like Higgs $H_1$ above the LEP bound of $114$~GeV.
\footnote{It is also possible to evade the LEP bound if $H_1$ decays
into the lightest pseudoscalars, with branching ratio
${\mathcal B}(H_1\to A_1A_1)>0.7$ \cite{NaturalRSymmetry}.
This requirement leads to a lower bound on the doublet component
of $A_1$, $\mathcal{O}^A_{11}>0.04$. Since the radiative corrections
are subdominant in this region, we do not include these points in our
results.}

The superpotential of the NMSSM exhibits a global $U(1)_R$ symmetry
which is spontaneously broken when $S$, the scalar component of
$\hat S$, acquires a VEV. In addition, it is explicitly broken by the
soft trilinear couplings $A_\lambda,A_\kappa$ \cite{Rsymmetry}.
The CP-odd scalar component of $\hat S$ is therefore
a pseudo-Goldstone boson of this symmetry, and is massless in the
limit $A_\lambda, A_\kappa \to 0$. For small values of the trilinear
couplings, the lightest pseudoscalar in the NMSSM spectrum can therefore
naturally be very light and highly gauge singlet in nature.
Typically this requires $A_\lambda \sim 200$~GeV, $A_\kappa \sim 5$~GeV.
Such a scenario can arise within the context of gauge- or gaugino-mediated
SUSY breaking, where both couplings are zero at tree level,
with non-zero $A_\lambda$ being radiatively generated at one loop 
and non-zero $A_\kappa$ at two loops \cite{NaturalRSymmetry}.

For a sufficiently light $A_1$ boson, observation in the decay $\Upsilon(1s)
\to \gamma A_1$ becomes a possibility.\footnote{In principle, this 
decay is also possible within other Higgs singlet extensions of the 
MSSM, such as the Minimal Non-minimal Supersymmetric extension of the
Standard Model (MNSSM)\cite{mnSSM}, and Eq.~(\ref{branchingratio}) 
is also valid in this case. However, the Higgs bosons of the MNSSM
obey a tree level mass sum rule, so that any light pseudoscalar boson
is accompanied by a quasi-degenerate scalar boson. Singlet-doublet 
mixing in the scalar sector typically excludes such a scenario except
in the MSSM limit of the theory $\lambda\to 0$ with $\mu_{\rm eff}$ fixed.
The radiative coupling of the singlet pseudoscalar to fermions, whose
effects we consider here, will also vanish in this limit.} Such a signal
has previously been considered in \cite{UpsilonDecay,Hiller}, with the
pseudoscalar coupling to $b$-quarks only through tree level singlet-doublet
mixing. It has recently been shown that the singlet Higgs bosons also receive
a direct coupling to fermions at one loop \cite{Hodgkinson}. Although loop
suppressed, this coupling is enhanced by the ratio of Higgs doublet VEVs
$\tan\beta$ and can become competitive with tree-level effects when this
parameter is large.

In this Letter we consider the effects of such a direct coupling on 
the decay $\Upsilon\to\gamma A_1$. Experimental searches
\cite{UnstableSearches} for a light Higgs boson in $\Upsilon$ decays
place a $90\%$ confidence level upper bound on the branching ratio
${\mathcal B}(\Upsilon \to \gamma A_1) \lsim 1\times 10^{-4}$
for a light particle $m_{A_1}<8$~GeV decaying visibly within the
detector. The upper bound rises to $\sim 10^{-3}$ for heavier particles
due to the softness of the recoil photon and cuts placed on energy deposits
in the detector tighten the constraints to $\sim 10^{-5}$ for a stable 
or invisibly decaying $A_1$ boson\cite{StableSearches}.

The branching ratio for $\Upsilon$ decays
through the Wilczek mechanism \cite{Hiller,WilczekMechanism} is given by

\begin{equation}
\label{branchingratio}
{{\mathcal B}(\Upsilon \to \gamma A_1)
\over {\mathcal B}(\Upsilon \to \mu^+\mu^-)}
=
{G_F m^2_\Upsilon \over 4\sqrt{2} \pi\alpha}
(g^P_{A_1 bb})^2
\left(1-{m_{A_1}^2\over m_\Upsilon^2}\right)F\ .
\end{equation}
Here $F\sim 1/2$ includes QCD corrections \cite{QCDFactor} and
${\mathcal B}(\Upsilon \to \mu^+\mu^-)=(2.48\pm0.06)\%$.
The SM-normalised pseudoscalar coupling $g^P_{A_i bb}$ is given by 
\cite{Hodgkinson}
\begin{equation}
\label{gpbb}
g^P_{A_i bb} =
\left(1+\frac{\sqrt{2}\left<\Delta_b\right>}{v_1}\right)^{-1}
\left[-\left(\tan\beta+\Delta_b^{a_2}\right){\mathcal O}^A_{1i}
+\Delta_b^{a_S} {{\mathcal O}^A_{2i}\over \cos\beta}\right],
\end{equation}
with ${\mathcal O}^A$ the $2\times2$ orthogonal pseudoscalar mixing matrix,
such that
\begin{equation}
A_1={\mathcal O}^A_{11} a+{\mathcal O}^A_{21} a_S,
\end{equation}
where $a$ is the would-be CP-odd scalar in the MSSM limit and $a_S$ is the
CP-odd singlet Higgs boson. In addition, $\Delta_b^{a_{2,S}}$ are the one-loop
non-holomorphic Yukawa couplings of the states $a_{2,S}$ to $b$ quarks. At
zero external momentum, they may be calculated by
\begin{equation}
\Delta_b^{a_{2,S}} = 
i\sqrt{2}\left<\frac{\partial\Delta_b[\Phi_1,\Phi_2,S]}
{\partial a_{2,S}}\right>,
\end{equation}
where $\Delta_b[\Phi_1,\Phi_2,S]$ is a Coleman-Wienberg type functional
\cite{CWFunctional} of the background Higgs fields which encodes radiative
corrections to the $b$ quark self-energy. Here $\left<\ldots\right>$ denotes
taking the VEV of the enclosed expression. The dominant contributions to
$\Delta_b^{a_{2,S}}$ are due to gluino-sbottom quark and chargino-stop quark
loops. In the single-Higgs-insertion approximation, neglecting subdominant
terms proportional to the weak gauge coupling $\alpha_w$, they may be given by

\begin{eqnarray}
\Delta_b^{a_2} & = & -\frac{2\alpha_S}{3\pi} \tilde M_3 \mu
I(\tilde M_Q^2,\tilde M_b^2,\tilde M_3^2)
-\frac{h_t^2}{16\pi^2} \mu A_t
I(\tilde M_Q^2,\tilde M_t^2,\mu^2)\ ,\\
\Delta_b^{a_S} & = & -\frac{2\alpha_S}{3\pi} \tilde M_3 \mu
\frac{v_2}{v_S}
I(\tilde M_Q^2,\tilde M_b^2,\tilde M_3^2)
-\frac{h_t^2}{16\pi^2} \mu A_t
\frac{v_2}{v_S}
I(\tilde M_Q^2,\tilde M_t^2,\mu^2)\ ,
\end{eqnarray}
where $\tilde M_{Q,t,b}$ are the soft squark masses, $A_t$ is the top-squark
soft trilinear coupling and $\tilde M_3$ is the gluino mass. The one-loop
function $I(x,y,z)$ is given by
\begin{equation}
I(x,y,z) =
\frac{xy\ln(x/y)+yz\ln(y/z)+xz\ln(z/x)}
{(x-y)(y-z)(x-z)}\ .
\end{equation}

\begin{figure}
\begin{center}
  \includegraphics{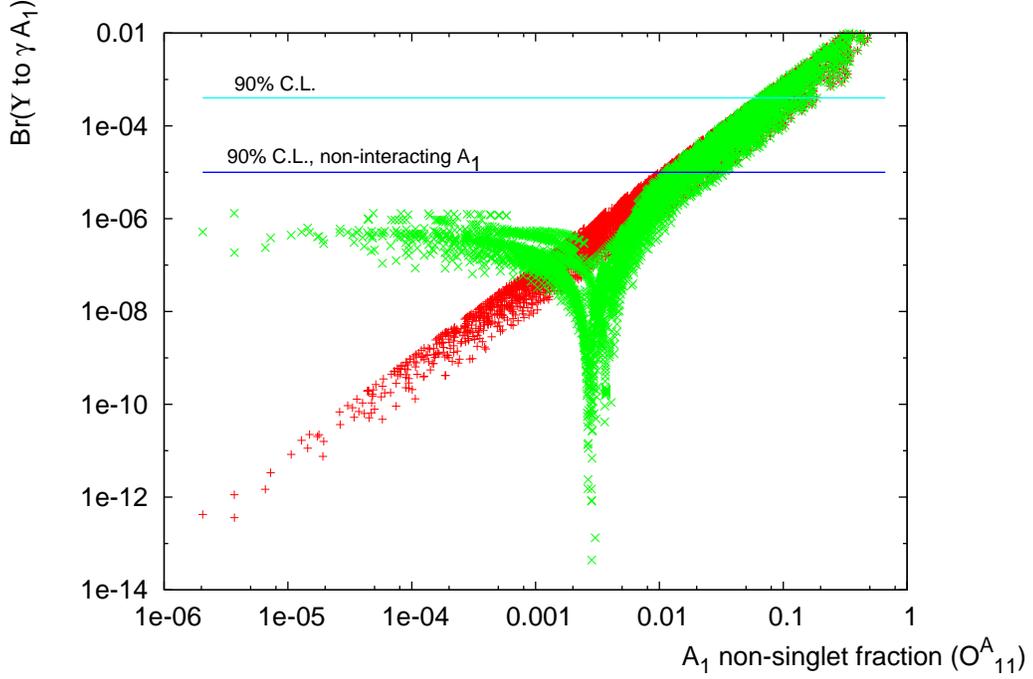}
\caption{\it The branching ratio ${\mathcal B}(\Upsilon \to \gamma A_1)$
vs. the non-singlet fraction $O^A_{11}$ at $\tan\beta=50$. The points in
green (light grey) include the one-loop threshold effects $\Delta_b^{a_s}$,
points in red (dark grey) neglect these corrections. Here
$\mu_{\rm eff}=120$~GeV and $\lambda, \kappa, A_\lambda, A_\kappa$ are
scanned over the range given in Eq.~(\ref{Scan}). All other soft-SUSY
breaking parameters are taken to equal $M_{\rm SUSY}=600$~GeV. Experimental
bounds are shown in dark blue (black) for a stable or invisibly decaying
pseudoscalar and in light blue (grey) for a visibly decaying particle,
assuming here $m_{A_1}\sim5$~GeV. The full limits are strongly dependent
on the value of $m_{A_1}$ and are less restrictive by one to two orders of
magnitude for a heavy $A_1$ boson ($m_{A_1}>8$~GeV).}
\label{results1}
\end{center}
\end{figure}

In Fig.~\ref{results1} we present results from a scan over the
parameters
\begin{equation}
\label{Scan}
\begin{array}{cc}
0<\lambda<0.5, & 0<A_\lambda<300 {\rm GeV},\\
-0.5<\kappa<0.5, &  0<A_\kappa<20 {\rm GeV},\nonumber
\end{array}
\end{equation} whilst fixing $\tan\beta=50$ and 
$\mu_{\rm eff}=120$~GeV. We require a light Higgs pseudoscalar  
$m_{A_1}<9$~GeV along with a lightest Higgs scalar $m_{H_1}>114$~GeV,
in agreement with constraints from LEP II. The soft-SUSY breaking 
parameters which enter the calculation of $\Delta_b^{a_{2,S}}$
are taken to be equal at $M_{\rm SUSY}=600$~GeV. The branching ratio 
${\mathcal B}(\Upsilon\to \gamma A_1)$ is plotted against
the non-singlet fraction of $A_1$, described by the mixing
matrix element ${\mathcal O}^A_{11}$. 

The threshold corrections are independent of the tree-level
coupling proportional to the pseudoscalar mixing, and enter
the expression for $g^P_{A_1bb}$ with opposing sign. For a 
relatively large non-singlet component above $few\ \%$, the 
threshold corrections represent a small suppression to the branching 
ratio of up to $\sim  10\%$. In the case of a highly singlet 
$A_1$ boson, the threshold corrections become the dominant effect,
producing a branching ratio of the order $\sim 1\times10^{-6}$ 
across a significant region of parameter space. This prediction is
found to be generic for electroweak-scale soft SUSY-breaking terms 
around a TeV. At the intersection of these regimes, the contributions 
cancel giving a highly suppressed decay rate.

\begin{figure}
\begin{center}
  \includegraphics{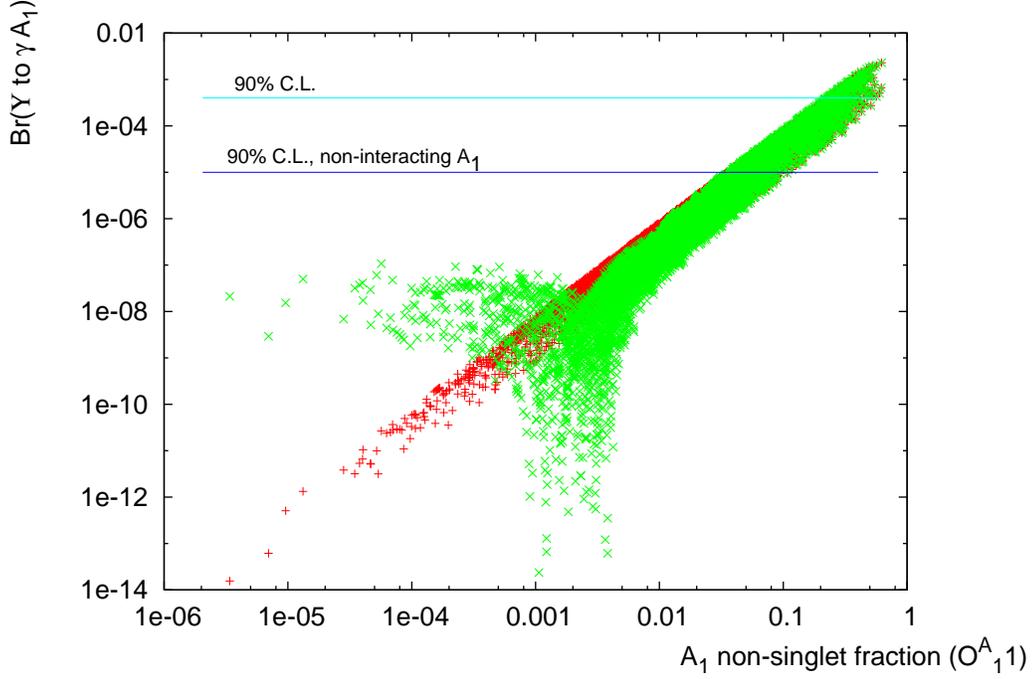}
\caption{\it The branching ratio ${\mathcal B}(\Upsilon \to \gamma A_1)$
vs. the non-singlet fraction $O^A_{11}$ at $\tan\beta=10$. Here
$\mu_{\rm eff}=120$~GeV and $M_{\rm SUSY}=600$~GeV, with $\lambda, \kappa,
A_\lambda, A_\kappa$ scanned over the range given in Eq.~(\ref{Scan}).
Points in green (light grey) include the one-loop threshold effects
$\Delta_b^{a_S}$, points in red (dark grey) neglect these corrections.
Experimental bounds are shown as for Fig.~\ref{results1}.}
\label{results2}
\end{center}
\end{figure}

Fig.~\ref{results2} shows results from a scan over the parameter range
of Eq.~(\ref{Scan}) for $\tan\beta=10$, keeping $\mu=120$~GeV and the
common soft-SUSY breaking scale $M_{\rm SUSY}=600$~GeV. Both the
doublet-singlet mixing and threshold correction contributions to
$g^P_{A_1 bb}$ are $\tan\beta$ enhanced, such that the branching
ratio at low $\tan\beta$ is smaller by $1\sim 2$ orders of magnitude
across the full parameter space. Due to their common enhancement, the
relative importance of the two terms in Eq.~$(\ref{gpbb})$ varies only
slowly with $\tan\beta$, so that for all values of $\tan\beta\gsim 5$,
minimal branching ratios are observed for singlet-doublet mixing around
$few\ \times 0.1\%$. The magnitude of the branching ratio is not found
to vary strongly with $M_{\rm SUSY}$ or $\mu$, although the available
parameter space consistent with out requirements $m_{H_1}>114$~GeV,
$m_{A_1}<9$~GeV decreases as $\mu$ increases, such that small values of
the singlet-doublet mixing $O^A_{11}$ do not appear.

The inclusion of threshold corrections can clearly alter the
phenomenology of highly singlet light pseudoscalars in a
dramatic way, allowing for the possibility of detectable 
$\Upsilon\to A_1\gamma$ decays in a new corner of parameter space.
In the limit of vanishing singlet-doublet mixing the tree level coupling
of the $A_1$ boson to $\tau$ leptons also vanishes, however an analogous
threshold correction also contributes to the $A_1 \tau^+\tau^-$ coupling
$g^P_{A_1\tau^+\tau^-}$, through a wino-stau loop. The pseudoscalar is
therefore predicted to decay into $\tau^+\tau^-$ pairs with branching ratio
of order one, for $2m_{\tau}<m_{A_1}<m_\Upsilon$, independently of the
singlet-doublet mixing. An order-of-magnitude estimate suggests that current
$B$-factories should be sensitive to branching ratios of the order
${\mathcal B}(\Upsilon \to \gamma A_1)\lsim 10^{-6}$, for observing such a
final state.

At masses above $\sim9$~GeV, the $A_1$ boson can mix with the $\eta_b$ meson.
This can lead to significant enhancement or suppression of 
${\mathcal B}(\Upsilon \to \gamma A_1)$\cite{ResonantMixing}. In addition,
there is a broadening of the $A_1$ width, and the resonance in the energy
spectrum of the recoil photon is less sharply peaked. There has been a
suggestion to search for such a light Higgs boson through precision tests of
lepton universality in the decays of the $\Upsilon$ \cite{LNU}. Such searches
would also be sensitive to decays in the zero-mixing limit. If the $A_1$ boson
is below the $\tau^+\tau^-$ threshold, the dominant decay channels are
$s\bar s, gg$ (and hence light mesons) or photon pairs, since the coupling to
$c\bar c$ is $\tan\beta$ suppressed.\footnote{The $\tan\beta$ suppression would 
also exclude the possibility of an observable signal $J/\psi\to \gamma A_1$ for
$m_{A_1}<2m_c$ in the limit of vanishing singlet-doublet mixing.} This
remains a favourable situation for the clean environment of an $e^+e^-$
collider, where these final states can be reliably measured.

Unfortunately, despite the tremendous production rates for
$b$-mesons at the LHC, a discovery of the $A_1$ boson through this mechanism
appears difficult. The final state consists of low-energy
$\tau$ jets and a photon, neither of which presents a clean signal above
background activity. An alternative production mechanism has been suggested
in \cite{AssociatedChargino}, which considers instead the process
$pp\to \tilde\chi_1^+\tilde\chi_1^- A_1$ in the limit of vanishing
doublet-singlet mixing. The possibility for observing such a signal
is strongly dependent on the masses and decay channels of both the
lightest chargino and the $A_1$ boson.

If both terms contributing to $g^P_{A_1 b\bar b}$ are of similar
magnitude, typically for around $\sim 0.5\%$ mixing, detection of
the $A_1$ boson may be extremely challenging. In this case, the
branching fraction of $\Upsilon\to A_1\gamma$ becomes extremely suppressed.
An alternative experimental strategy is to look for $A_1$ pair production
from Higgs boson decays \cite{HtoAAsignals}. In the small singlet-doublet
mixing scenario at large $\tan\beta$, the lightest CP-even Higgs boson $H_1$
is highly SM-like, and the branching fraction ${\mathcal B}(H_1 \to A_1 A_1)$
is conservatively bounded from above at around $\sim 10^{-3}$. Associated
production of the $A_1$ boson with a chargino pair would remain a possibility.

In conclusion, we have shown that the branching ratio for
production of a light Higgs pseudoscalar in $\Upsilon(1s)$
decays does not vanish in the absence of doublet-singlet mixing.
We found that the decay $\Upsilon\to\gamma A_1$ is predicted to be
observable at existing experimental facilities if supersymmetry is
broken at the TeV scale with large $\tan\beta$. In the event of a
cancellation between the threshold corrections and tree-level mixing
contributions to the $A_1 b\bar b$ coupling the branching ratio may be
highly suppressed even though the doublet-singlet mixing is still
significant, and further phenomenological considerations would be needed.
We hope to return to this issue in a future communication.

\subsection*{Acknowledgments}

The author would like to thank Apostolos Pilaftsis and Roger Barlow for
helpful discussions. This research was supported by the U.K. Science and
Technology Facilities Council.




\end{document}